\documentclass[11pt, letterpaper]{article}

\usepackage{amsmath}
\usepackage{amssymb}
\usepackage{amsthm}
\usepackage{graphicx}
\usepackage{hyperref}
\usepackage{url}
\usepackage{algorithm}
\usepackage{algorithmic}
\usepackage{caption}
\usepackage{subcaption}
\usepackage{booktabs}
\usepackage{authblk}
\usepackage{appendix}
\usepackage[utf8]{inputenc}
\usepackage[T1]{fontenc}
\usepackage[english]{babel}
\usepackage{geometry}
\usepackage{xcolor}
\usepackage{fancyhdr}
\usepackage{microtype}
\usepackage{parskip}
\usepackage{tabularx}
\usepackage{enumitem}

\hypersetup{
    colorlinks=true,
    linkcolor=blue,
    filecolor=magenta,      
    urlcolor=cyan,
    citecolor=red,
}

\definecolor{headercolor}{RGB}{0, 50, 100}
\title{Evaluating Large Language Models on a Highly-specialized Topic, Radiation Oncology Physics}

\author[1]{Jason Holmes}
\author[2]{Zhengliang Liu}
\author[1]{Lian Zhang}
\author[1]{Yuzhen Ding}
\author[1]{Terence T. Sio}
\author[1]{Lisa A. McGee}
\author[1]{Jonathan B. Ashman}
\author[3]{Xiang Li}
\author[2]{Tianming Liu}
\author[1]{Jiajian Shen\thanks{*co-corresponding author: shen.jiajian@mayo.edu}}
\author[1]{Wei Liu\thanks{*co-corresponding author: liu.wei@mayo.edu}}

\affil[1]{Department of Radiation Oncology, Mayo Clinic}
\affil[2]{School of Computing, The University of Georgia}
\affil[3]{Department of Radiology, Massachusetts General Hospital}

\date{}

\begin{document}

\maketitle

\begin{abstract}
\noindent We present the first study to investigate Large Language Models (LLMs) in answering radiation oncology physics questions. Because popular exams like AP Physics, LSAT, and GRE have large test-taker populations and ample test preparation resources in circulation, they may not allow for accurately assessing the true potential of LLMs. This paper proposes evaluating LLMs on a highly-specialized topic, radiation oncology physics, which may be more pertinent to scientific and medical communities in addition to being a valuable benchmark of LLMs. We developed an exam consisting of 100 radiation oncology physics questions based on our expertise at Mayo Clinic. Four LLMs, ChatGPT (GPT-3.5), ChatGPT (GPT-4), Bard (LaMDA), and BLOOMZ, were evaluated against medical physicists and non-experts. ChatGPT (GPT-4) outperformed all other LLMs as well as medical physicists, on average. The performance of ChatGPT (GPT-4) was further improved when prompted to explain first, then answer. ChatGPT (GPT-3.5 and GPT-4) showed a high level of consistency in its answer choices across a number of trials, whether correct or incorrect, a characteristic that was not observed in the human test groups. In evaluating ChatGPT's (GPT-4) deductive reasoning ability using a novel approach (substituting the correct answer with "None of the above choices is the correct answer."), ChatGPT (GPT-4) demonstrated surprising accuracy, suggesting the potential presence of an emergent ability. Finally, although ChatGPT (GPT-4) performed well overall, its intrinsic properties did not allow for further improvement when scoring based on a majority vote across trials. In contrast, a team of medical physicists were able to greatly outperform ChatGPT (GPT-4) using a majority vote. This study suggests a great potential for LLMs to work alongside radiation oncology experts as highly knowledgeable assistants.
\end{abstract}

\section{Introduction}
\label{sec:introduction}
The advent of large language models (LLM) has completely transformed natural language processing (NLP) ~\cite{zhao2023brain}. The traditional paradigm of NLP follows the typical pipeline of creating customized solutions for downstream applications through supervised training. For example, a pre-trained BERT ~\cite{devlin2018bert} model must be modified with additional network layers and then fine-tuned on labeled training data to perform tasks such as sequence classification or question answering. In some situations, it might also be beneficial or necessary to further pre-train such models on domain specific data to attain acceptable performance ~\cite{gu2021domain,liu2022survey}. For example, AgriBERT ~\cite{rezayi2022agribert} was pre-trained on agriculture-related text data, to properly address NLP tasks in the food and agriculture domain. However, the expansive size and exceptional few-shot learning capabilities enable LLMs to solve NLP problems through \textit{in-context learning}, which reduces or even eliminates the need for annotated training samples ~\cite{brown2020language,dong2022survey}. During in-context learning, LLMs generalize from a few examples (or no examples at all) based on \textit{prompts}, which typically are descriptive user inputs that characterize desired responses from LLMs ~\cite{brown2020language,liu2023pre}. For example, "summarize the following text" is a straightforward prompt that asks the language model to produce a summary for the input text. In general, LLMs provides a novel and simplified workflow for NLP that could potentially do away with supervised fine-tuning and its associated intricacies such as hyper-parameter tuning and model architecture modification. Furthermore, in-context learning significantly reduces the need for expensive and time-consuming human annotation ~\cite{brown2020language,dai2023chataug}. It is especially desirable in medicine and science due to the limited data available in these domains ~\cite{kagawa2022one,rezayi2022clinicalradiobert,liu2022survey,liu2023deid}. 

In recent months, the world has witnessed the rise of of ChatGPT \footnote{https://openai.com/blog/chatgpt}, which has enjoyed significant global popularity given its unprecedented language capabilities and accessibility to the general public through a chatbot interface. ChatGPT is based on the powerful GPT-3 model ~\cite{brown2020language}, one of the first large language models in history. The 175-billion-parameters GPT-3 was trained on a large data collection that encapsulated diverse Internet data (including the Common Crawl \footnote{http://commoncrawl.org/} and Wikipedia \footnote{https://www.wikipedia.org/}). It demonstrates exceptional performance in a variety of NLP tasks spanning from text summarization to named entity recognition (NER) through its text generation objective (indeed, many NLP tasks can be translated to some forms of text generation). ChatGPT inherits these capabilities from GPT-3, along with the massive knowledge on diverse topics stored in the parameter space. More importantly, ChatGPT was trained through \textit{Reinforcement Learning from Human Feedback} (RLHF), a reinforcement learning process that incorporates human preferences and human ranked values through user feedback. This process tunes the model to generate outputs that are most appealing and relevant to human users. The capabilities of ChatGPT empowers diverse practical applications ranging from essay writing to code generation ~\cite{qin2023chatgpt}.

One of the most powerful LLM to date is GPT-4 \footnote{https://openai.com/research/gpt-4}, a successor to GPT-3. While OpenAI has not revealed much about its technical details yet, GPT-4 has demonstrated superior performance over the GPT-3.5-based ChatGPT in various scenarios ~\cite{openai2023gpt4,dai2023chataug,liu2023deid,koubaa2023gpt}. In fact, as of March 2023, GPT-4 is powering Microsoft's search engine, Bing ~\cite{BingGPT4}, which demonstrates the potential of LLM-based search. In addition, unlike its predecessors, GPT-4 is a multi-modal model that accepts image inputs, which undoubtedly leads to more interesting applications. 

GPT-4 has been shown to perform exceptionally well on various academic and professional benchmarks ~\cite{openai2023gpt4}. For example, GPT-4 passes the USMLE exam with a $>$20\% margin \cite{nori2023capabilities}. In fact, GPT-4 scores at over the 90th percentile on the SAT, the Uniform Bar Exam and the verbal section of the GRE (see Figure 4 in the "GPT-4 Technical Report" \cite{openai2023gpt4}), where almost all of them included a multiple-choice component. Indeed, multiple-choice examinations are common for evaluating LLMs~\cite{lampinen2022can,openai2023gpt4,savelka2023large}. Most multiple-choice exams that have been used to evaluate LLMs are based on topics that are among the most well represented in academics. For example, in 2022, the AP physics exam had 144,526 test-takers \cite{collegeboard2022}, the LSAT had 128,893 test-takers \cite{lsac2023}, the GRE had approximately 342,000 test-takers \cite{gre2023}. As a result of the large numbers of test-takers taking these exams as well as the importance placed on scores in determining university admittance, there exists an exceeding amount of resources (including text data accessible on the internet). Regardless of the particular LLM under evaluation, the ease of access and overall ubiquity of these tests and relevant test preparation materials effectively preclude a high performance when evaluating LLMs on these tests. It is therefore important to also study LLMs on more obscure and specialized topics where the size of the training data is likely much smaller. In 2022, there were only 162 medical school graduates, who applied for radiation oncology residency programs \cite{radmatch2023}. Radiation oncology physics therefore represents a topic that is relatively unknown to the general population and may therefore be a more fair test in evaluating LLMs as compared to highly represented knowledge-bases. Obscure topics may represent the greatest educational opportunity and also the greatest risk for the general population in the context of LLMs, as the responses may be more relied upon while being less accurate and with mistakes being less likely to be noticed.


An important factor in evaluating the accuracy of LLMs is to ensure that the test questions are left out of the training data \cite{bubeck2023sparks}, i.e.\,not contaminated. The best way to ensure this is to create new questions for testing. In this study, a multiple-choice examination has been created for this purpose. Four transformer-based LLMs have been chosen for evaluation: ChatGPT (GPT-3.5) \cite{brown2020language}, ChatGPT (GPT-4) \cite{openai2023gpt4}, Bard (LaMDA) \cite{thoppilan2022lamda}, and BLOOMZ \cite{muennighoff2022crosslingual}. These results are compared to radiation oncology experts as well as non-experts. Additionally, ChatGPT (GPT-4) is further explored on how to improve its answers and on its deductive reasoning capabilities. 

\section{Related Work}
\label{sec:relatedwork}

\subsection{Large language models}
Transformer-based pre-trained language models (PLMs), such as BERT \cite{devlin2018bert} and GPT \cite{radford2018improving}, have revolutionized natural language processing. Surpassing previous methods (e.g., RNN-based models) in numerous tasks, they have promoted interest in and accessibility of language models \cite{kalyan2021ammus}. Generally, PLMs can be categorized into three types: autoregressive models (like GPT), masked language models (such as BERT), and encoder-decoder models (e.g., BART \cite{lewis2019bart} and T5 \cite{raffel2020exploring}). More recently, there is a rise of very large language models, including GPT-3 \cite{brown2020language}, Bloom \cite{scao2022bloom}, PaLM \cite{chowdhery2022palm}, and OPT \cite{zhang2022opt}. Rooted in the transformer architecture, these models draw inspiration from the likes of BERT and GPT but are developed at much larger scales.

The objective of large language models is to accurately learn contextual and domain-specific latent feature representations from input text \cite{kalyan2021ammus}. For example, the vector representation of "discharge" might vary considerably between medical and general domains. Smaller language programs often require continual pre-training and supervised fine-tuning on downstream tasks to achieve acceptable performance \cite{gu2021domain,liu2022survey}. However, very large language models could potentially eliminate the need for fine-tuning while maintaining competitive results \cite{brown2020language}.

Besides the progress in model architecture, scale and training strategies, large language models can be further aligned with human preferences through reinforcement learning from human feedback (RLHF) \cite{ziegler2019fine}. This approach has been implemented in various LLMs, such as Sparrow \cite{glaese2022improving}, InstructGPT \cite{ouyang2022training}, and ChatGPT. InstructGPT was based on GPT-3 and was trained through a process during which user preferences were prioritized through human-generated ranking feedback. As a successor to InstructGPT, ChatGPT also employs RLHF, focusing on adhering to prompts and generating comprehensive responses. OpenAI also implemented guardrails to prevent the generation of biased and undesirable outputs \cite{chatgpt2023}. ChatGPT has become a highly successful AI chatbot, capitalizing on GPT-3.5's capabilities to facilitate human-like interactions.

RLHF incorporates human feedback into the generation and selection of optimal results by training a reward model based on human annotators' rankings of generated outcomes \cite{bai2022training}. This reward model then rewards outputs that best correspond to human preferences and values. We believe these groundbreaking innovations make ChatGPT the perfect candidate for this study.

The recent development of GPT-4 has significantly advanced the state-of-the-art of language models. GPT-4 demonstrates enhanced reasoning abilities, creativity, image comprehension, context understanding, and multi-modal abilities, leading to more sophisticated and diverse responses. The success of large GPT models spurs exploration into specialized variants for specific fields, such as dedicated large language models for medical and healthcare applications, which could potentially revolutionize these domains.

\subsection{Language models and examination}
Large language models have exceptional natural language comprehension abilities. In addition, they are trained on massive data that supplies substantial knowledge. These characteristics make large language models ideal candidates for academic and professional benchmarks.

OpenAI recently released the first study in the literature that evaluates large language models on academic and professional exams designed for educated humans \cite{openai2023gpt4}. The results indicate that GPT-4 performs extremely well on a wide variety of subjects ranging from the Uniform Bar Exam to GRE. In addition, a study from Microsoft indicates that GPT-4 can pass USMLE, the professional exam for medical residents, by a large margin \cite{nori2023capabilities}. 

This study is the first evaluation of large language models in the realms of radiation oncology and medical physics, and we believe it can inspire future research in evaluating LLMs on highly-specialized branches of medicine.

\subsection{Prompt engineering}
Collecting and labeling data for training or fine-tuning NLP models can be resource-intensive and costly, especially in the medical domain \cite{liu2022survey,dai2023chataug,liu2023deid}. Recent studies suggest that by employing prompts, large-scale pre-trained language models (PLMs) can be adapted to downstream tasks without the need for fine-tuning \cite{brown2020language,liu2023pre}.

A prompt consists of a set of instructions that customizes or refines the LLM's response. Prompts extend beyond merely describing the task or specifying output formats. Indeed, they can be engineered to create novel interactions. For example, it is possible to prompt ChatGPT to emulate a cybersecurity breach with simulated terminal commands \cite{white2023prompt}. In addition, prompts can also be used to generate additional prompts through a self-adaptation process \cite{white2023prompt}. 

The emergence of prompt engineering signifies the start of a new era in natural language processing \cite{liu2023pre}. There is no doubt that carefully crafted prompts have much potential for diverse and sophisticated applications. However, determining the ideal prompt poses a unique challenge in the age of large language models. Currently, prompts can be designed manually or generated automatically \cite{gao2020making,liu2023pre}. Although automatically produced prompts may outperform manual prompts in certain tasks \cite{liu2023pre}, they often suffer from poor human-readability and explainability \cite{taylor2022clinical,liu2023pre}. Consequently, manual prompt generation may be favored in domains where interpretability is crucial, such as clinical practices and research. In this study, we design a suite of prompts and chain-of-thought prompts based on our experience in radiation oncology and medical physics and evaluate their impact on large language models. 

\section{Methods}
\label{sec:methods}
A 100-question multiple-choice examination on radiation oncology physics was created for this study by an experienced medical physicist, designed to match the curriculum for radiation oncology resident education as recommended by the American Society for Radiation Oncology (ASTRO) \cite{PMID:27354135}. This exam includes questions on the following topics: basic physics (12 questions), radiation measurements (10 questions), treatment planning (20 questions), imaging modalities and applications in radiotherapy (17 questions), brachytherapy (13 questions), advanced treatment planning and special procedures (16 questions), and safety, quality assurance (QA), and radiation protection (12 questions). Of the 100 questions, 17 require numeric calculation (math-based). The exam questions are listed in the Appendix, Section \ref{appendix:test_questions}.

\subsection{Comparison between LLM scores and human scores}
\label{subsec:methods_comparison}
The 100-question multiple-choice test on radiation oncology physics was inputted to each LLM in 5 separate trials (Trial 1 - Trial 5), except BLOOMZ, which was only tested in one trial. Each trial, beginning on a new thread or after reset, began with an initialization prompt notifying the LLM that it was about to be tested. Next, the LLM was prompted with instructions and 20 questions in batches until the exam was complete. In each trial, the instructions indicated to the LLM that it should only return the correct answer with no justification. The instructions were included in each batch since it was observed that the LLMs were less likely to follow the instructions otherwise. In cases where the LLM could not accept 20 questions at a time, batches of 10 questions were used instead (Bard). In cases where not all the answers were returned by the LLM, the next batch would include the non-answered question(s) as well as the entire next batch. These occurrences were rare. In each test trial, the global prompt and instructions prompt were phrased differently in order to account for response-noise due to prompt-noise. The initialization prompts and instructions prompts are given in Table \ref{tab:LLM_prompts}.

\begin{table}
    \centering
    \caption{The LLM prompts used in each trial.}
    \label{tab:LLM_prompts}
    \begin{tabularx}{\textwidth}{lXX}
        \toprule
        Trial & Initialization prompt & Instructions prompt \\
        \midrule
        Trial 1 & I am a radiation therapy researcher. My research group would like to study the answers given by ChatGPT on the topic of radiation oncology physics. I will now proceed to ask questions about radiation oncology physics. & Instructions: For each multiple choice question, provide the correct answer without any justification. \\
        Trial 2 & I want to evaluate your knowledge on radiation oncology physics by asking some multiple choice questions. & Please give only the question label and the letter for the correct answer. \\
        Trial 3 & Please answer the following practice questions as if you were a resident preparing for board certification exams. & Only give the correct answer in your response. Do not explain your answers. \\
        Trial 4 & We want to test your understanding of radiation oncology physics. For this reason, we have created some questions to ask you. & In your response, only report the question label and the corresponding answer. \\
        Trial 5 & I will ask you some multiple-choice questions. & Instructions: Only respond with the correct letter choice. \\
        \bottomrule
    \end{tabularx}
\end{table}

LLM test scores and their distributions were compared between each other as well as with scores from two human groups, medical physicists and non-experts. The medical physicists group included four experienced board-certified medical physicists, three medical physics residents, and two medical physics research fellows. The non-expert group included six individuals with advanced degrees in either electrical engineering, computer engineering, or computer science, but with no known prior experience or education on radiation oncology physics. Each human test-taker was allowed 3 hours to take the exam, closed book, also permitting the use of a basic calculator. In comparing the LLM scores and human scores, the mean scores, consistency in scores, and confidence in answers were evaluated.

To quantify the overall consistency of scoring success, the standard deviation and average correlation between trials, defined as the average of the upper values of the Pearson correlation matrix between trials, were calculated. The average correlation indicates how consistent the correct scores were between trials where 1 is interpreted as the distribution being identical, 0 is equivalent to the distribution being purely random, and -1 is interpreted as the distribution being perfectly anti-correlated.

In order to quantify the degree of confidence in the answers given by the LLMs and human groups, the number of correct answers for each question were counted across all trials. For example, if each LLM answered the same question correctly 5 times, then the percentage of questions where all 5 answers were correct was incremented by 1\% (since there are 100 questions). Additionally, the test results were compared to the expected distribution that would occur if the test-taker were guessing at random. The expected number of correct answers in 5 trials, when randomly guessing, is approximately $0.25 \times 5 = 1.25$ on average (98/100 questions have 4 choices, 2/100 have 5 choices). Using this value, the number of correct answer occurrences for each question can be estimated following the resultant Poisson distribution.

Finally, ChatGPT (GPT-3.5 and GPT-4) and Bard scores were compared to human scores where the scores were calculated based on majority vote.

\subsection{Improving ChatGPT (GPT-4) accuracy - explain first, then answer}
\label{subsec:methods_accuracy}
Due to the nature of transformer-based LLMs predicting the next word based on the prior context, it has been shown that the accuracy of responses can be improved if a sufficiently large LLM is prompted to develop the answer in a step-wise manner \cite{bubeck2023sparks, shinn2023reflexion, wei2022emergent}. ChatGPT (GPT-4) was evaluated using this strategy to see if its score could be improved by prompting it to explain first, then answer. The initialization prompt was the same as in Trial 1, however the instructions prompt for Trial 1 was changed to the following: "Instructions: For each multiple choice question, first give an explanation for the answer followed by the correct answer (letter choice)." These test results were then compared with the original non-justified test results.

\subsection{Testing ChatGPT (GPT-4) on its deductive reasoning ability}
\label{subsec:methods_deductive}
In a multiple-choice question, an LLM will be most successful when the question and answer are often used in the same context. However, what happens if the correct answer has no shared context with the question, such as when the answer is "None of the above"? In this case, the LLM must deduce the correct answer by rejecting all the other answers, all of which likely share context with the question. This scenario would seem to be especially challenging for an LLM. To study the deductive reasoning ability of ChatGPT (GPT-4), each question of the 100-question multiple-choice exam was modified. Each correct answer was removed and replaced with "None of the above choices is the correct answer." Such a context-reduction transformation cannot be used on a human since a human would notice the pattern. Because of this, there are likely to be no examples of this sort of transformation to be found for tests that were designed for humans and were subsequently used in the training data for LLMs. It is assumed, then, that an LLM would not notice this pattern. The modified exam was given to ChatGPT (GPT-4) using the Trial 1 prompts and was subsequently tested for improving accuracy by explaining first, then answering as described in Section \ref{subsec:methods_accuracy}.

\section{Results}
\label{sec:results}

\subsection{Comparison between LLM scores and human scores}
\label{subsec:results_comparison}
\begin{figure}
    \centering
    \includegraphics[width=1.0\textwidth]{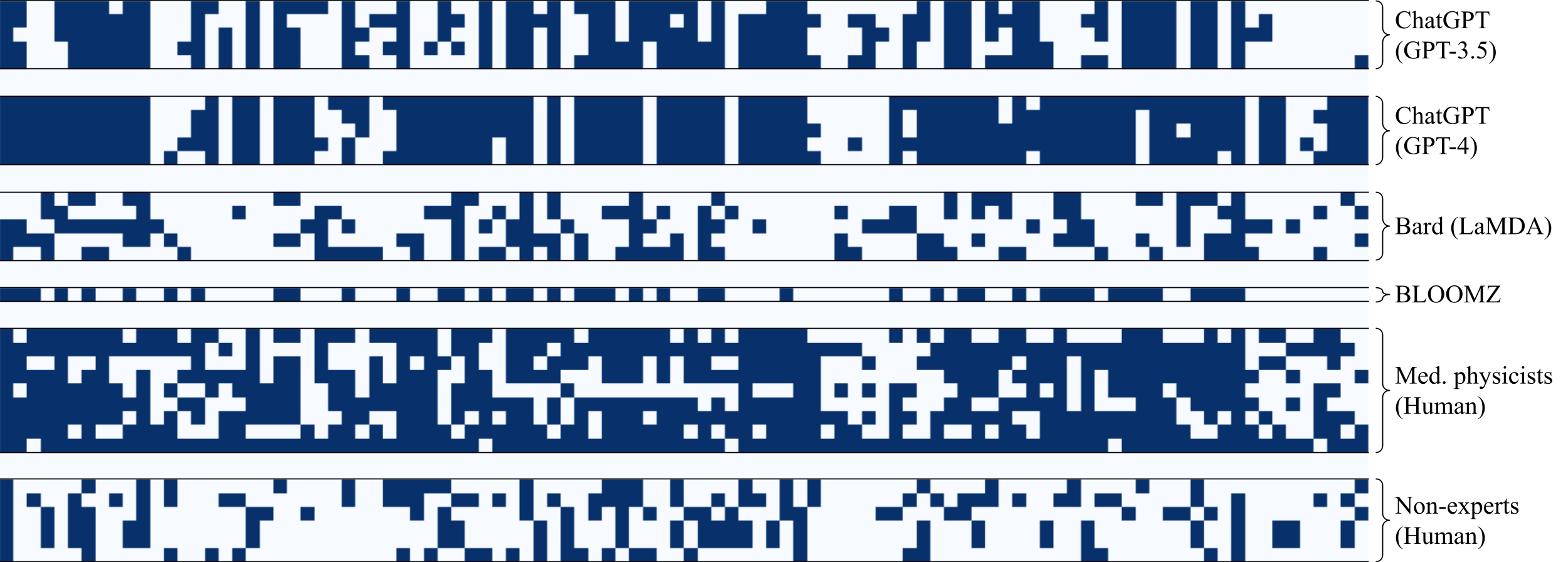}
    \caption{Raw marks for each test where the rows are separate tests and the columns are the test questions. Dark shaded squares represent correct answers.}
    \label{fig:test_scores_raw}
\end{figure}

\begin{figure}
    \centering
    \includegraphics[width=1.0\textwidth]{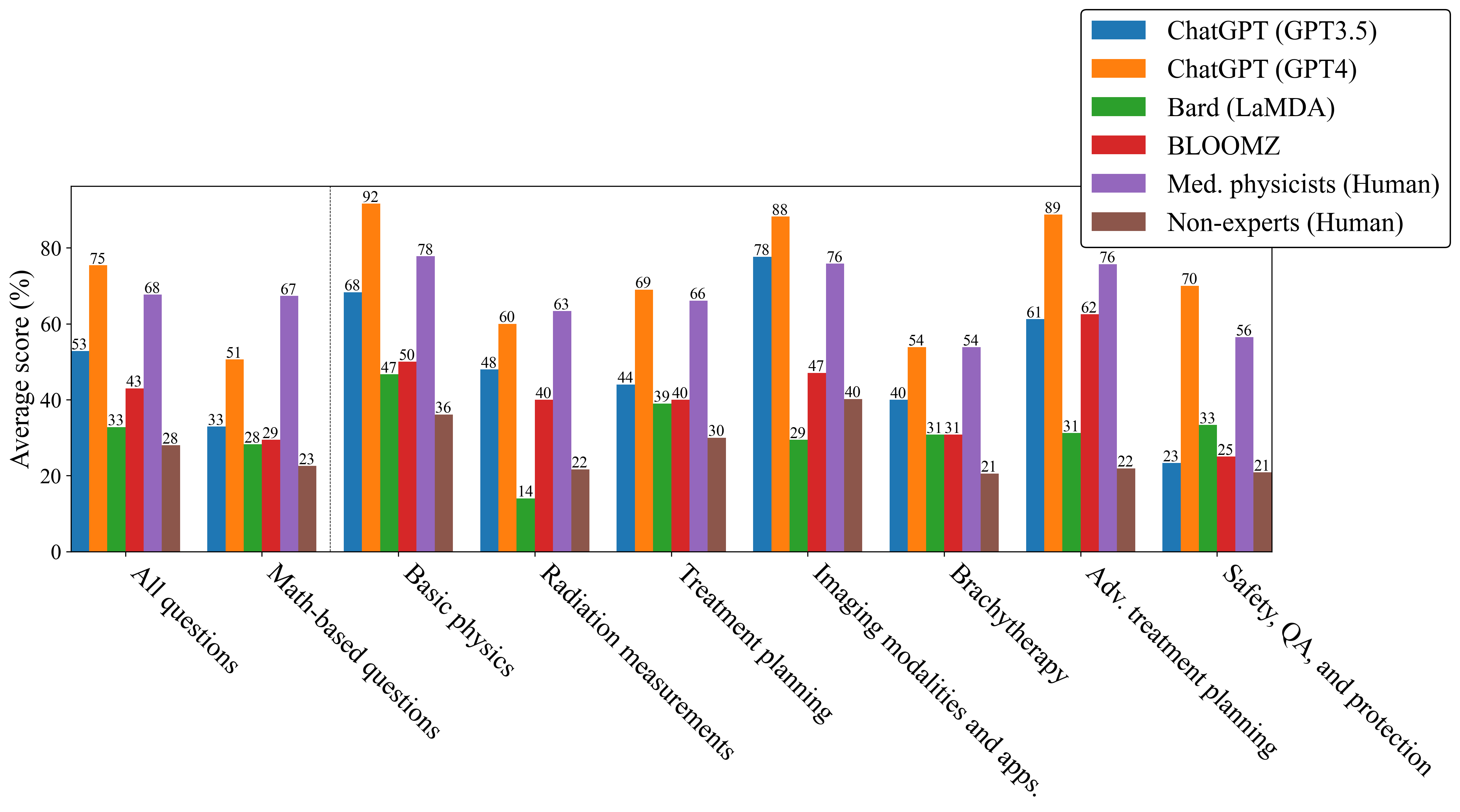}
    \caption{Mean test scores for each LLM by category.}
    \label{fig:test_scores}
\end{figure}

The raw marks and mean test scores are shown in Figures \ref{fig:test_scores_raw} and \ref{fig:test_scores} respectively, where the LLM mean test scores represent the mean of 5 trials (except for BLOOMZ - 1 trial) and the mean test scores for humans represent the mean of their respective groups (see Section \ref{subsec:methods_comparison}). Each LLM was able to outperform the non-expert human group overall while only ChatGPT (GPT-4) outperformed the medical physicist group. For math-based questions, the medical physicists outperformed ChatGPT (GPT-4).

\begin{figure}[htbp]
  \centering
  \begin{subfigure}[b]{0.49\textwidth}
    \includegraphics[width=\textwidth]{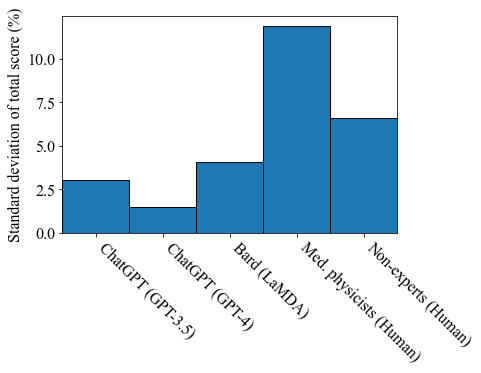}
    \caption{Standard deviation in total scores.}
    \label{fig:standard_deviation}
  \end{subfigure}
  \hfill
  \begin{subfigure}[b]{0.49\textwidth}
    \includegraphics[width=\textwidth]{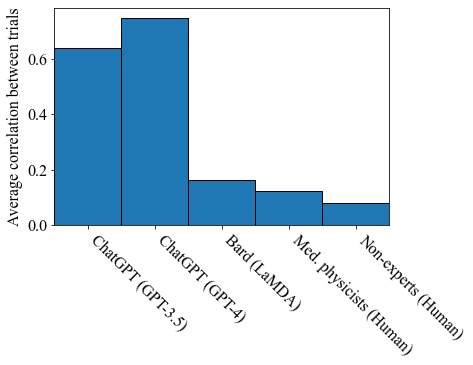}
    \caption{Average correlation between trials.}
    \label{fig:average_correlation}
  \end{subfigure}
  \caption{Consistency in scoring for LLMs and humans.}
  \label{fig:consistency}
\end{figure}

 As can be observed in the raw marks shown in Figure \ref{fig:test_scores_raw}, each LLM and human group showed variability between trials, not only in terms of uncertainty in the overall score, but also in terms of the number of times each question was answered correctly. The standard deviation and average correlation between trials are reported in Figures \ref{fig:standard_deviation} and \ref{fig:average_correlation}. The LLMs were much more consistent in their scores and answers as compared to the human groups, showing both a low standard deviation in scoring and a high average correlation between trials.

\begin{figure}[htbp]
    \centering
    \begin{subfigure}[b]{0.45\textwidth}
        \centering
        \includegraphics[width=\textwidth]{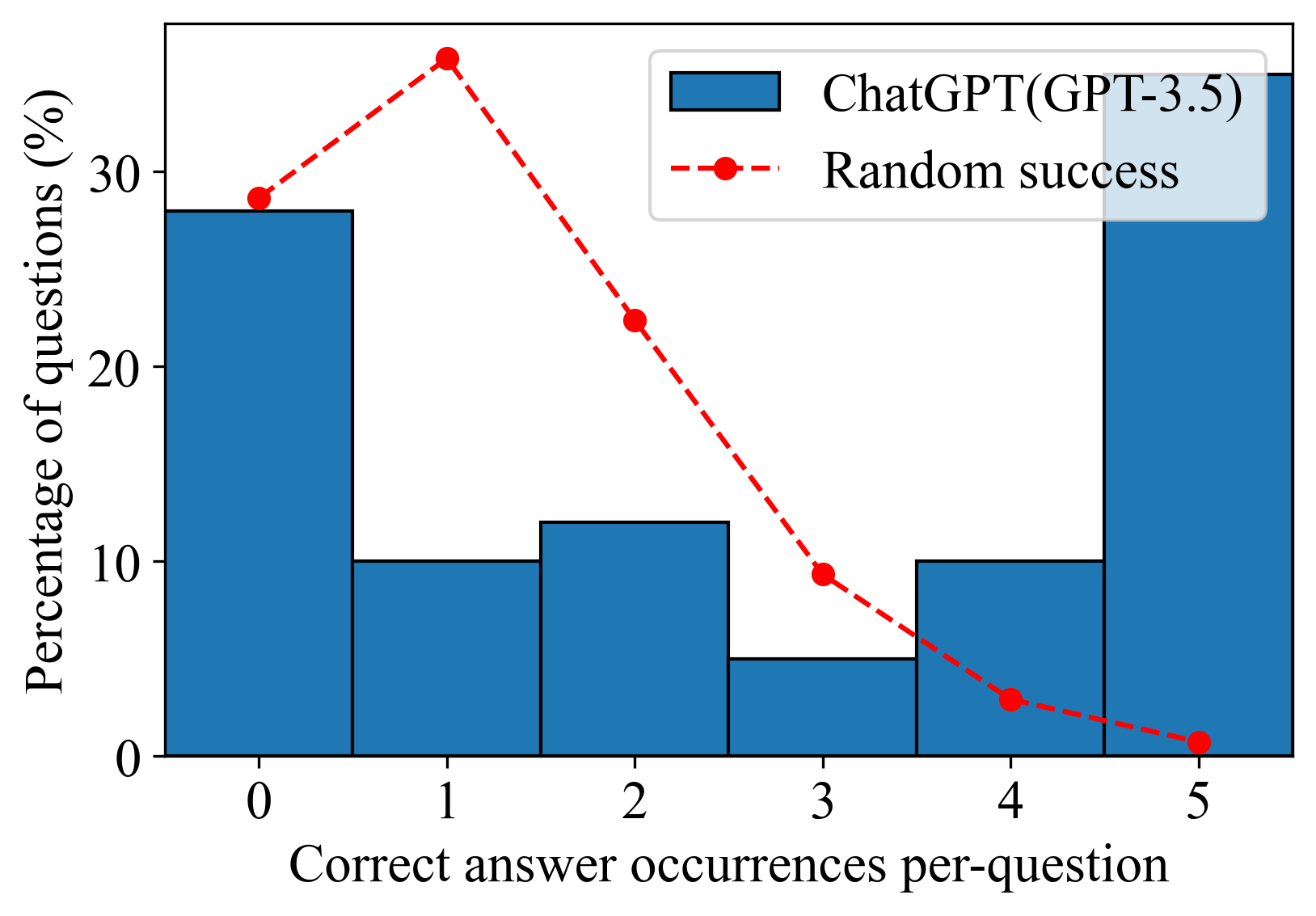}
        \caption{}
        \label{fig:dist_gpt3}
    \end{subfigure}
    \hfill
    \begin{subfigure}[b]{0.45\textwidth}
        \centering
        \includegraphics[width=\textwidth]{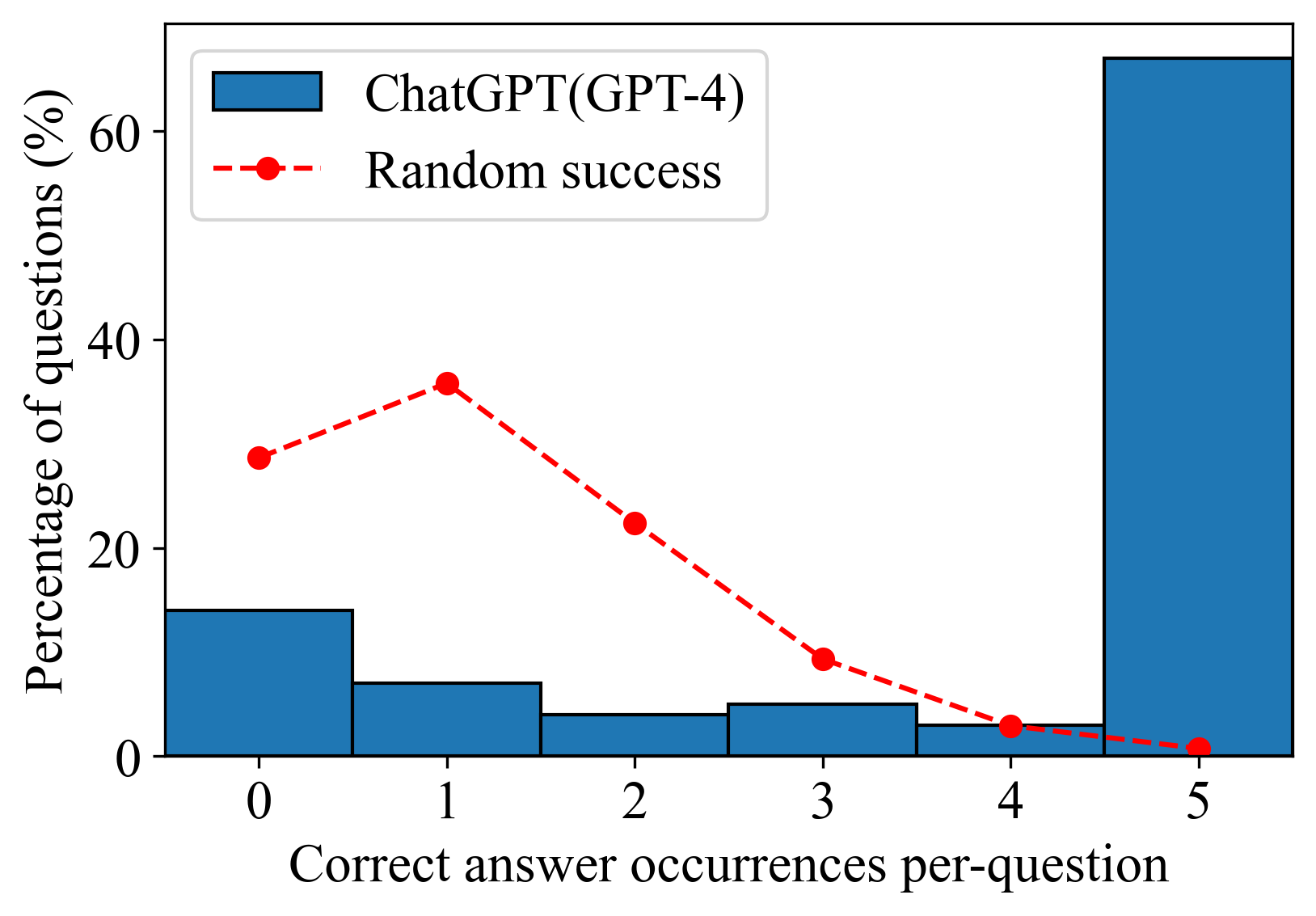}
        \caption{}
        \label{fig:dist_gpt4}
    \end{subfigure}
    
    \vspace{1cm}
    
    \begin{subfigure}[b]{0.45\textwidth}
        \centering
        \includegraphics[width=\textwidth]{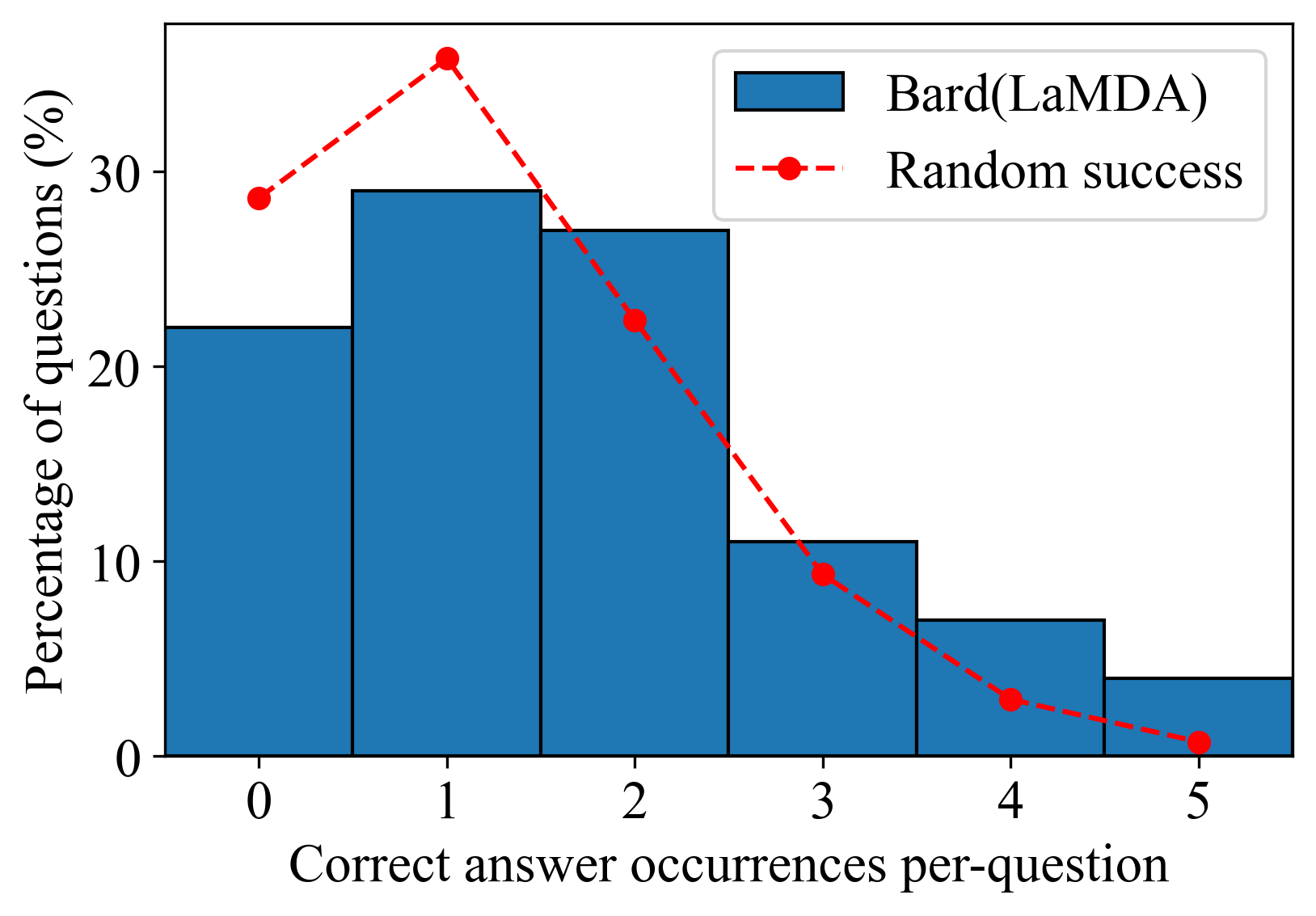}
        \caption{}
        \label{fig:dist_bard}
    \end{subfigure}

    \vspace{1cm}

    \begin{subfigure}[b]{0.45\textwidth}
        \centering
        \includegraphics[width=\textwidth]{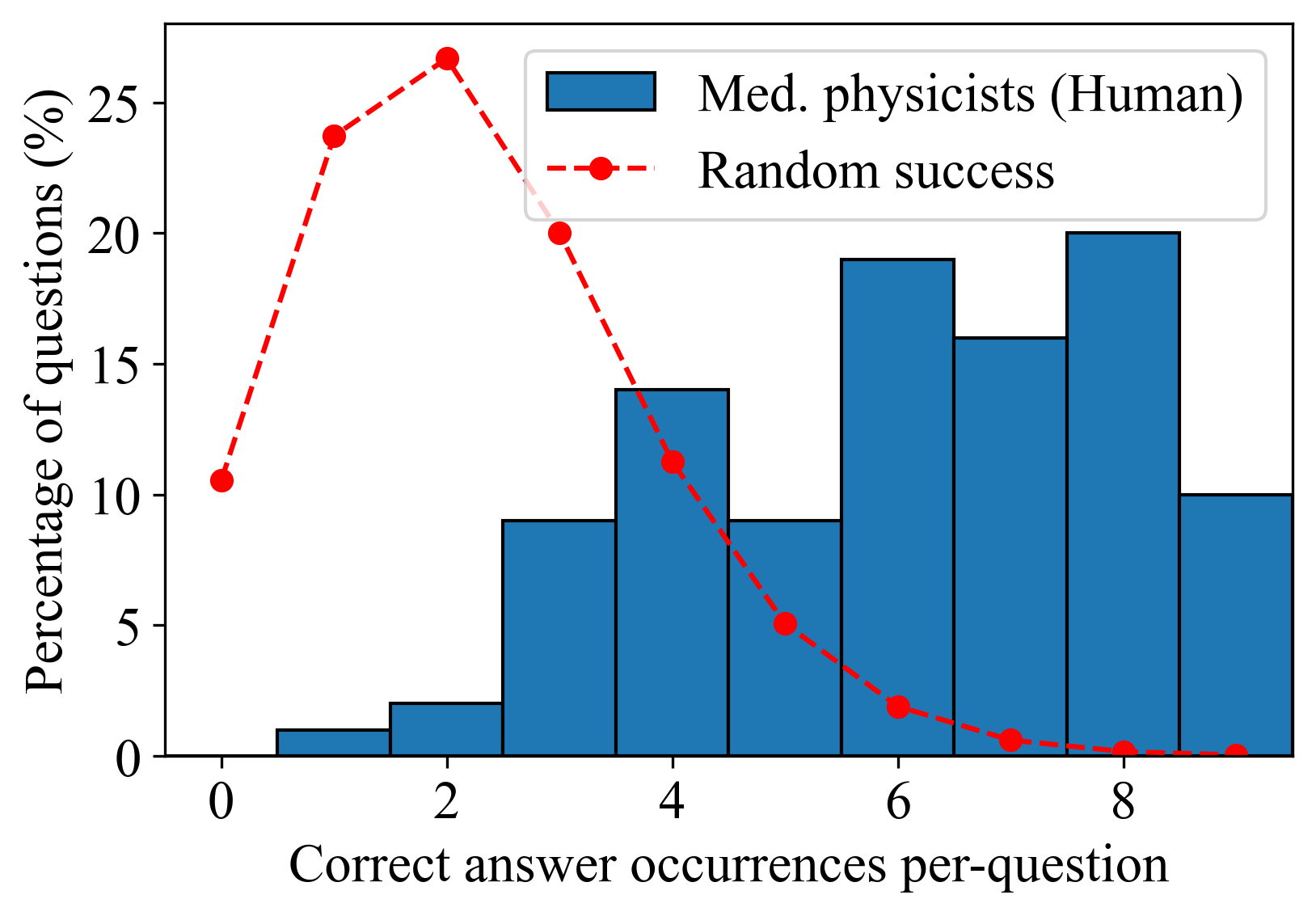}
        \caption{}
        \label{fig:dist_phys}
    \end{subfigure}
    \hfill
    \begin{subfigure}[b]{0.45\textwidth}
        \centering
        \includegraphics[width=\textwidth]{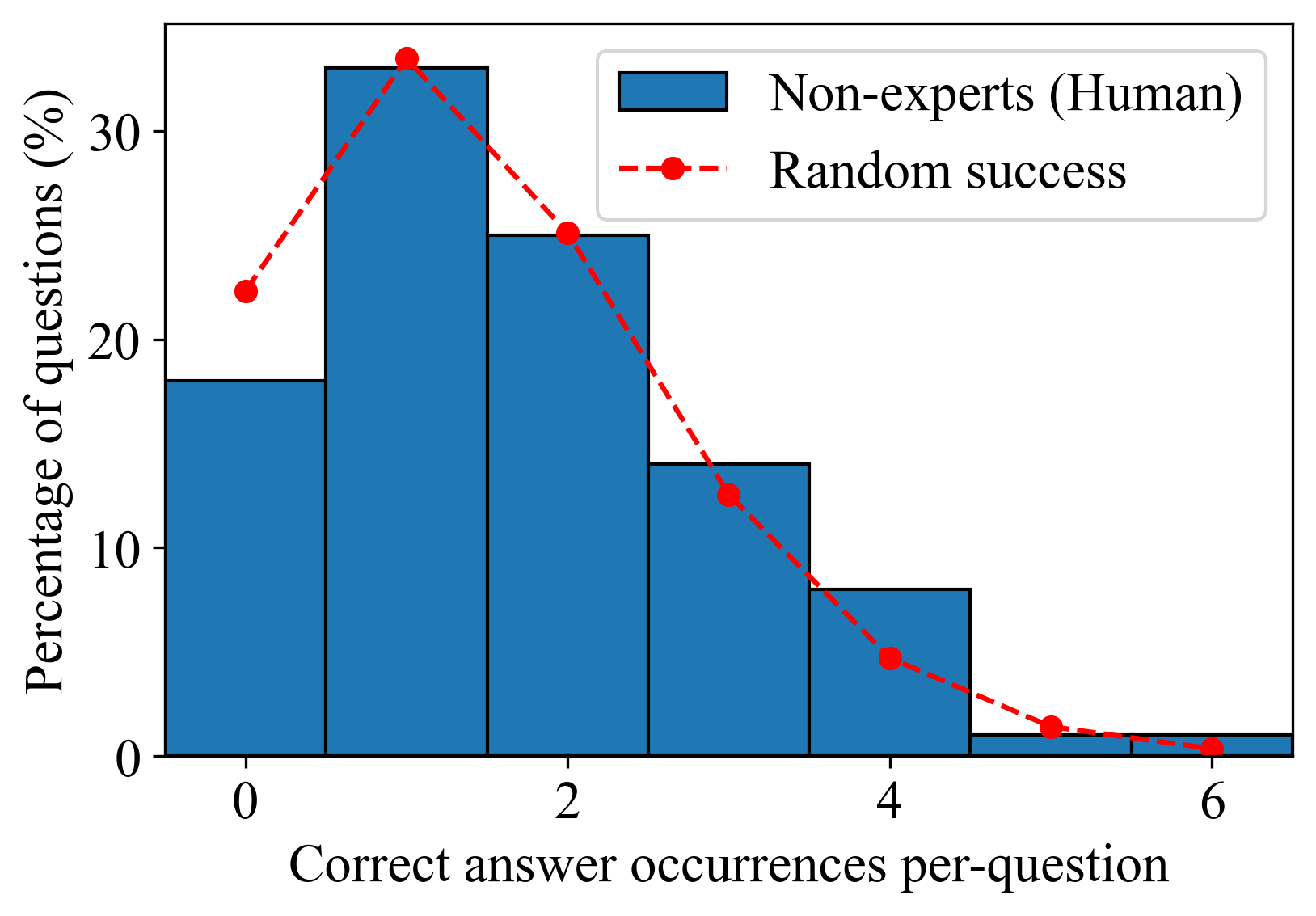}
        \caption{}
        \label{fig:dist_non}
    \end{subfigure}

    \caption{Confidence in answers. The number of correct answer occurences per-question for each LLM and human group. The dashed red curve indicates the expected distribution if the answers were randomly selected based on the Poisson distribution.}
    \label{fig:score_dist}
\end{figure}

From the results shown in \ref{fig:score_dist}, Bard slightly outperformed the non-expert group, however both groups performed similarly to a random guesser. ChatGPT (GPT-3.5 and GPT-4) and the medical physicists showed no similarity to random guessing. ChatGPT (GPT-3.5) was either confident, getting 35\% of answers correct in each trial, or confused, getting 28\% of answers incorrect. ChatGPT (GPT-4) was even more confident, getting 67\% of questions correct in each trial, however it also showed a propensity for confusion, getting 14\% of questions incorrect in each trial. As a group, the medical physicists were neither extremely confident, nor confused, however tending towards agreement in selecting the correct answers.

\begin{figure}
    \centering
    \includegraphics[width=1.0\textwidth]{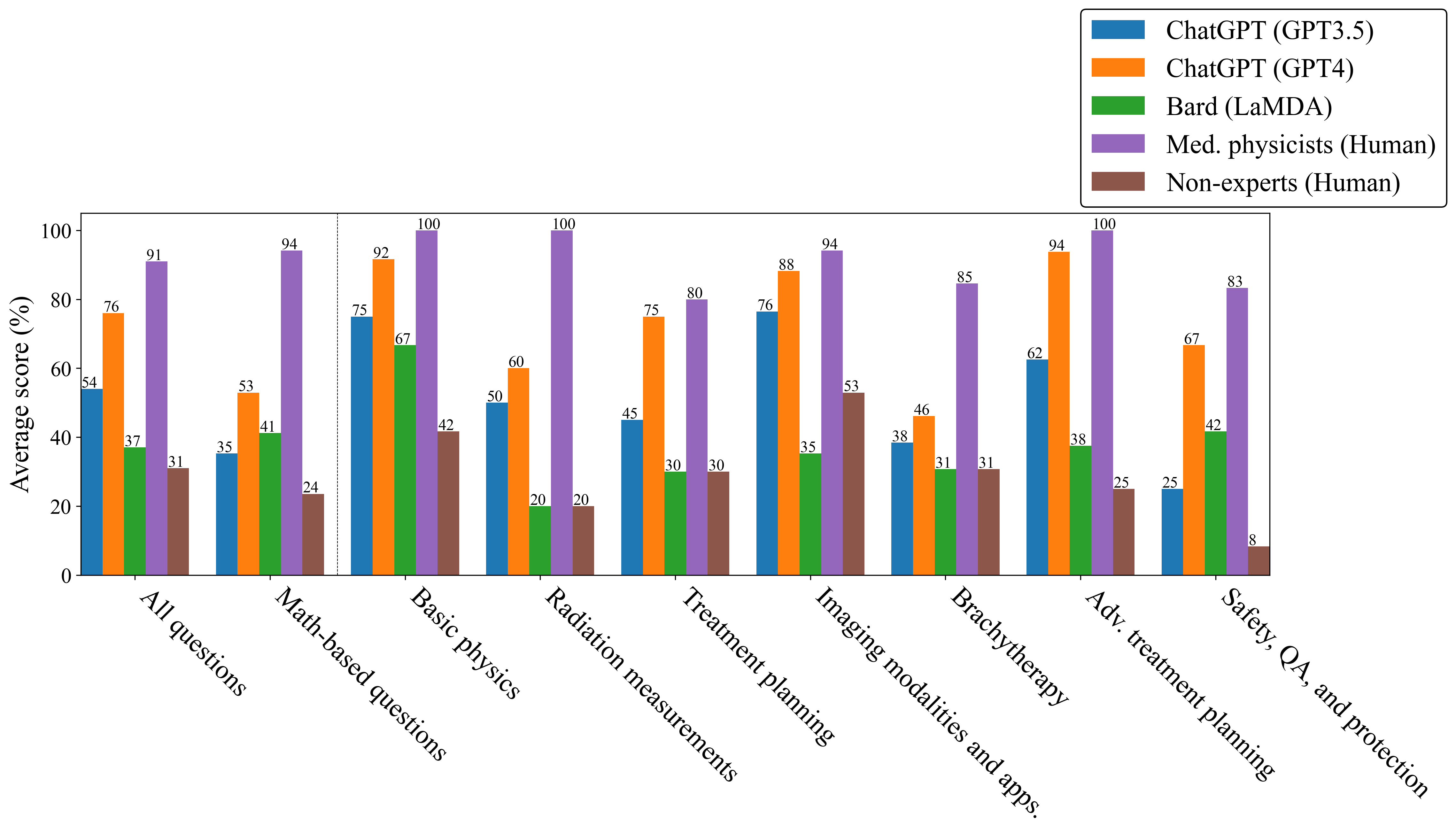}
    \caption{Scores by category, tabulated by majority vote among trials for LLMs and within the group for humans.}
    \label{fig:majority_vote}
\end{figure}

Although ChatGPT (GPT-3.5 and GPT-4) scored well overall, their scoring distributions, shown in Figure \ref{fig:score_dist}, suggested that if the LLMs could work together, there would be very little improvement in scoring, since they tended to be either confident or confused with low variability. Bard (LaMDA) and the non-expert groups would also likely show little improvement in working together as their answers tended towards random success. However, because medical physicists tended towards agreement on correct answers, it would be expected that their score would improve considerably when working together. To test for this, the answers for each group were combined using a "majority vote". For each question, the most common answer choice was chosen as the group answer. In the case of a tie, one answer among the most common answer choices was chosen randomly. Figure \ref{fig:majority_vote} shows the scoring results when utilizing a majority vote. As shown, ChatGPT (GPT-3.5 and GPT-4) improved very slightly, 1\%. Bard (LaMDA) and the non-expert group improved by 4\% and 3\% respectively. However, the medical physicist group improved greatly, by 23\%.

\subsection{Improving ChatGPT (GPT-4) accuracy - explain first, then answer}
\label{subsec:results_accuracy}
\begin{figure}
    \centering
    \includegraphics[width=0.63\textwidth]{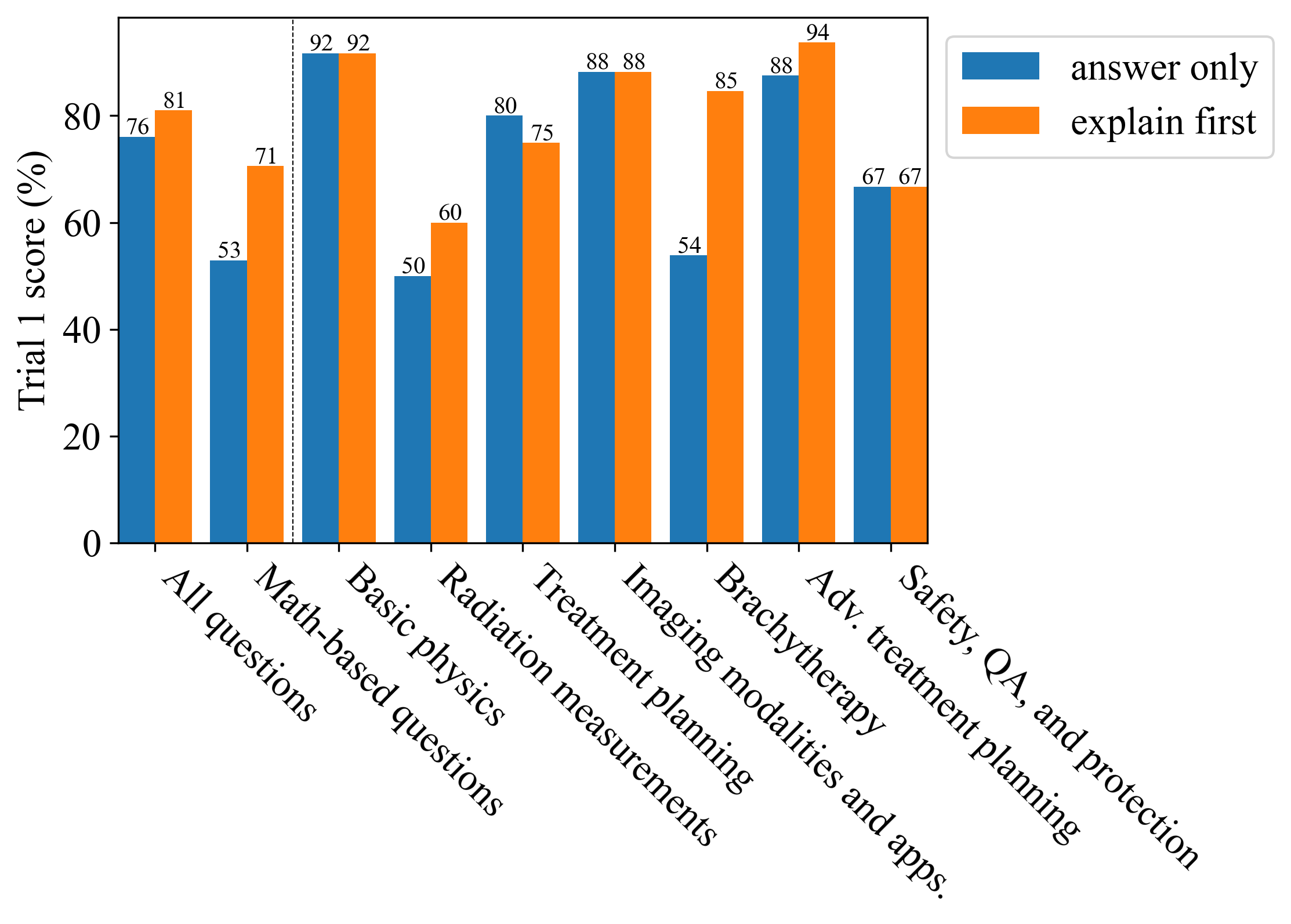}
    \caption{The improvement for Trial 1 as due to using the explain first, then answer method.}
    \label{fig:explain_first}
\end{figure}

Figure \ref{fig:explain_first} shows the results for having prompted ChatGPT (GPT-4) to explain first, then answer, therefore allowing the answer to develop. ChatGPT's (GPT-4) overall score improved by 5\%, exceeding each prior trial. The greatest improvement was in the brachytherapy and math-based questions categories. These results are in agreement with prior studies that found this capability to be an emergent characteristic for sufficiently large LLMs ~\cite{wei2022emergent}. Sample responses from ChatGPT (GPT-4) are given in the Appendix, Section \ref{appendix:explainfirst}.

\subsection{Testing ChatGPT (GPT-4) on its deductive reasoning ability}
\label{subsec:results_deductive}
\begin{figure}
    \centering
    \includegraphics[width=0.63\textwidth]{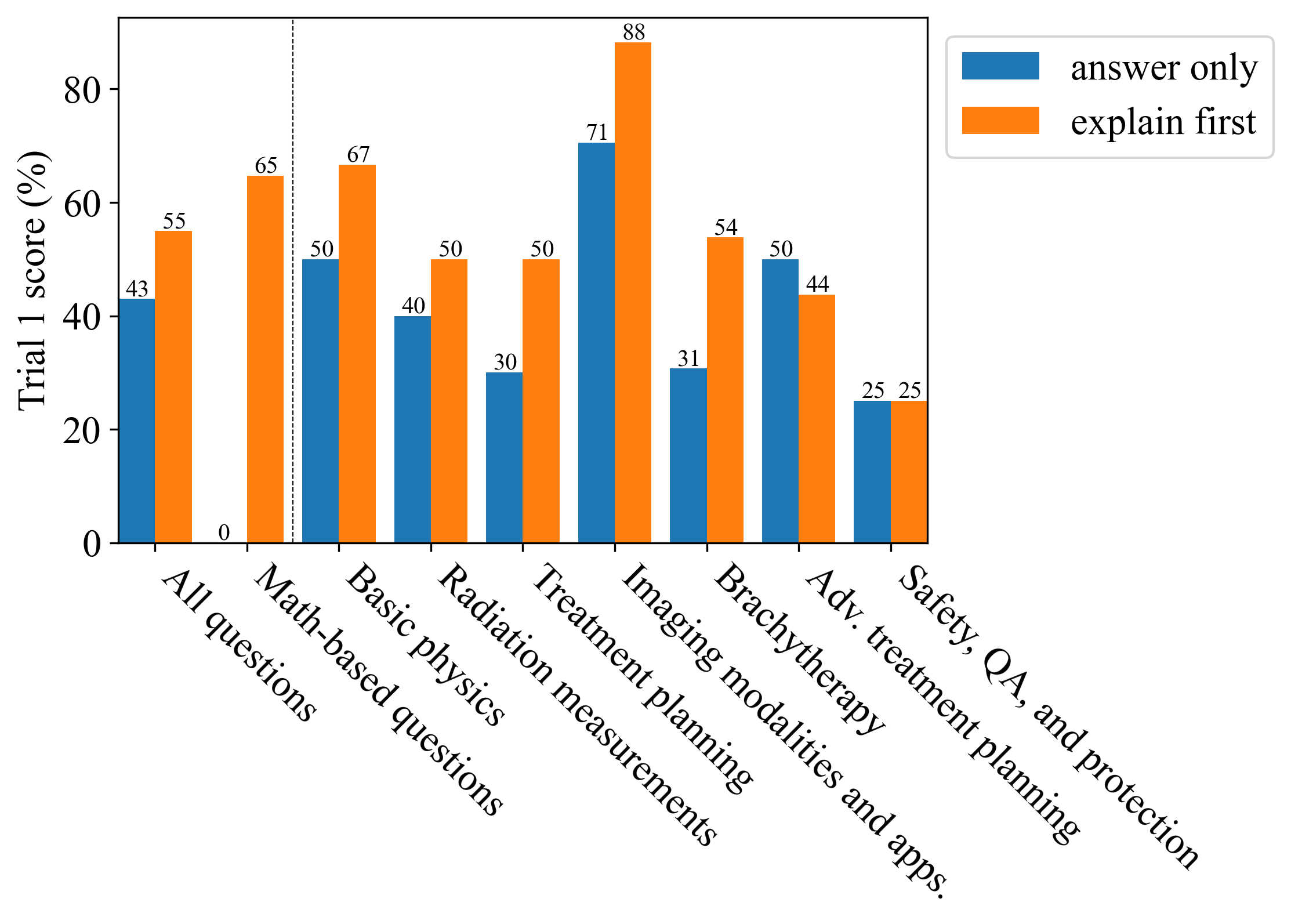}
    \caption{The scores for Trial 1 after replacing the correct answer with "None of the above choices is the correct answer.", a method for testing for deductive reasoning, and subsequent improvement as due to using the explain first, then answer method.}
    \label{fig:dt_explain_first}
\end{figure}

Figure \ref{fig:dt_explain_first} shows the results for the deductive reasoning test where the correct answer was replaced by "None of the above choices is the correct answer" in all 100 questions. Overall, ChatGPT (GPT-4) performed much more poorly as compared to the original test. Although the performance was generally worse, the explain first, then answer method was especially important in improving its ability to deductively reason through the questions. Without explaining first, ChatGPT (GPT-4) got 0\% of math-based questions correct, which improved to 65\% after incorporating the explain first, then answer method, only one question less accurate than the original trial also using the explain first, then answer method.

\section{Discussion}
\label{sec:discussion}
The aim of this study was to evaluate LLMs on a highly-specialized topic, radiation oncology physics, based on a 100-question multiple choice exam that was specifically designed for this study. The exam can be found in the Appendix, Section \ref{appendix:test_questions}. The scoring results from the non-expert group suggest that the general population knows very little about radiation oncology physics as their scores were similar to random guessing. Bard (LaMDA) slightly outperformed the non-experts while BLOOMZ and ChatGPT (GPT-3.5 and GPT-4) greatly outperformed the non-experts. Amazingly, GPT-4 was able to outperform the average medical physicist in nearly all subcategories and improved its answer accuracy when prompted to explain its reasoning before answering (Figures \ref{fig:test_scores} and \ref{fig:explain_first}). ChatGPT (GPT-4) showed a surprising ability to deductively reason in answering all 100 questions where each correct answer was modified to be "None of the above choices is the correct answer.", particularly when it was prompted to explain first, then answer, scoring 55\% overall. This result is somewhat perplexing and could potentially be an emergent property. Emergent properties are known to occur as the number of parameters is increased in LLMs \cite{wei2022emergent}. This novel method may be a useful method in determining whether deductive reasoning improves with the number of parameters going forward.

While ChatGPT (GPT-4) outperformed medical physicists overall, this study has also provided evidence that individual LLMs cannot compete with a small number of medical physicists working together (Figure \ref{fig:majority_vote}). The likely reason is that humans vary significantly in capabilities and knowledge from individual to individual, even when their professional backgrounds are similar. Additionally, while an answer in a multiple-choice question will either be correct or incorrect, the scoring count distributions shown in Figure \ref{fig:score_dist} indicated that the medical physicists were far less likely to be confused, which, when aggregated over the whole group of medical physicists, allowed them to select the correct answer at a much higher rate in a majority vote. When ChatGPT (GPT-3.5 and GPT-4) was wrong, it was confidently wrong (confused). Similarly, when it was correct, it was confidently correct. Our results indicated that humans with expertise on a highly-specialized topic knew when to guess, how to guess intelligently, and were less likely to be wrong in their reasoning, even when the correct answer was not chosen. This comparison may not be completely fair as it is possible that if the exact same human could be tested repeatedly in the same manner as ChatGPT (GPT-3.5 and GPT-4), they might also repeat answers and show a degree of confusion individually. That point is arguably irrelevant, however, as there are many experienced medical physicists and only few LLMs as capable as GPT-4. The high degree of consistency in correct and incorrect answers for ChatGPT (GPT-3.5 and GPT-4) may be a sign of over-fitting (or memorization) in regards to radiation oncology physics knowledge. Regardless, being that radiation oncology physics is a highly-specialized topic, the performance of ChatGPT (GPT-4) was extraordinary and will likely continue to improve in the near-future. Practically speaking, this study suggests a great potential for radiation oncology experts to work alongside ChatGPT (GPT-4), using it as a highly knowledgeable assistant.

A weakness in evaluating LLMs using exams such as the one presented in this study is that this exam is not representative of the detailed and nuanced daily clinical work being performed by medical physicists and radiation oncology specialists. The relative performance between LLMs and medical physicists on radiation oncology physics exams reported in this study may therefore misrepresent the degree of equivalency between LLMs and individual medical physicists. Furthermore, GPT-4's high performance on this certification-like exam, covering a highly specialized topic, suggests a degree of superficiality in the knowledge being assessed. Otherwise, we would have to entertain the possibility of GPT-4 being competent enough to fulfill the role of a medical physicist, which seems highly improbable. The radiation oncology community, and possibly the wider medical community, may therefore need to reevaluate certification procedures, as the necessity for humans to invest significant effort in acquiring such superficial knowledge will diminish as LLMs continue to advance. With this in mind, LLMs could potentially be used as a test for superficiality. Perhaps a greater focus on knowledge not known by the LLM should be more greatly emphasized.

\subsection{Applying large language models in radiation oncology}
This study is a continuation of a line of research that applies state-of-the-art NLP methods to radiation oncology. For example, Rezayi et al. \cite{rezayi2022clinicalradiobert} trained BioBERT on a large corpus of radiation oncology literature and a curated and anonymized text dataset from a hospital to build ClinicalRadioBERT, a specialized language model for radiation oncology. Liao et al. \cite{liao2023mask} proposed a framework of directing the attention of transformer-based language models to more important input tokens that significantly affect classification decisions. This method is particularly important for few-shot learning with few annotated samples, which is a common challenge in radiation oncology where it is difficult to collect and curate large amounts of multi-institution patient data that match certain requirements due to the concern of patient privacy. On a related note, ChatGPT has demonstrated superior performance as an effective text data augmentation approach over state-of-the-art text data augmentation methods in terms of testing accuracy and distribution of the augmented samples ~\cite{dai2023chataug}, which can also be used to address the few-shot learning challenge. 

In addition, LLMs can be employed for innovative applications such as data de-identification. For example, GPT-4 outperforms ChatGPT and other language model competitors in de-identifying clinical notes with a 99\% accuracy ~\cite{liu2023deid}. This is of extreme importance to radiation oncology and all medicine specialities in general, since it is often cumbersome to anonymize data for cross-institution clinical collaboration and academic research. Some other applications of language models include building domain-specific knowledge graphs for oncology \cite{cai2022coarse} without manual annotation from clinicians or other domain experts.

\subsection{Multi-modal models in radiation oncology}
Multi-modal models are the future of language models \cite{zhao2023brain} and are important in medical diagnosis \cite{cai2021chestxraybert}. Some early LLM studies with multi-modal data include ChatCAD \cite{wang2023chatcad}, a framework to integrate images and texts for computer-aided diagnosis. It supports various diagnosis networks such as those for lesion segmentation and report generation. In this framework, ChatGPT can be used to enhance the outputs of these networks. 

GPT-4 supports multi-modal inputs such as images, which further unlocks the potential of large language models in radiation oncology. It is necessary to investigate future models and applications that integrate text, images, dosimetric data, and other modalities into the diagnosis and treatment pipelines. We believe such multi-modal models display inherent affinity to the human brain ~\cite{zhao2023brain} and future LLM models for medicine can receive inspirations from advances in both neuroscience and NLP. 

\section{Conclusion}
This study demonstrated that LLMs, particularly ChatGPT (GPT-4), can excel in answering questions on a highly-specialized topic like radiation oncology physics as well as exhibit surprising capabilities in deductive reasoning. While GPT-4 outperformed individual medical physicists on average, a small group of medical physicists working together still achieved better results, demonstrating the importance of diversity in knowledge and expertise among human professionals - and perhaps for LLMs as well. This further highlights the potential benefits of combining LLMs' capabilities with human insight and experience to enhance knowledge and decision-making. Overall, the findings suggest a promising future for LLMs in specialized fields.

\section*{Acknowledgments}
We would like to thank the many individuals (Yuzhen Ding, Yiran Cui, Yonghui Fan, Yikang Li, Riti Paul, Nupur Thakur, Sachin Chhabra, Chenbin Liu, Yang Li, Man Zhao, Yuchao Hu, Shumei Jia, Lian Zhang, Yao Xu, Hongying Feng, and Yunze Yang) that volunteered to take the radiation oncology physics exam.

\newpage
\bibliographystyle{unsrt}
\bibliography{LLM_refs}

\newpage
\appendix

{\huge Appendix}
\section{Explain first, then answer responses from ChatGPT (GPT-4)}
\label{appendix:explainfirst}
In order to show how the explain first, then answer prompt works in ChatGPT (GPT-4), we selected 10 questions here for demonstration, starting from the 61$^{\text{st}}$ question of the 100-question found in the Appendix, Section \ref{appendix:test_questions}.
\begin{enumerate}[start=61] 
\item 1 Curie (Ci) is equal to $3.7 \times 10^{10}$ Becquerels (Bq). Therefore, 10 Ci of $^{192}$Ir is equivalent to $37 \times 10^{10}$ Bq or 370 GBq.\\
Answer: C. Its decay rate is 37 GBq.

\item The equation for effective half-life ($T_\text{eff}$) is $\frac{1}{T_\text{eff}} = \frac{1}{T_\text{phys}} + \frac{1}{T_\text{bio}}$, where $T_\text{phys}$ is the physical half-life and $T_\text{bio}$ is the biological half-life. Solving for $T_\text{bio}$ gives $T_\text{bio} = \left(\frac{1}{T_\text{eff}} - \frac{1}{T_\text{phys}}\right)^{-1} = \left(\frac{1}{2} - \frac{1}{6}\right)^{-1} = 3\,\text{hours}$.\\
Answer: D. 3 hours

\item The initial dose rate ($D_0$) can be calculated using the formula $D_0 = \frac{D_\text{total}}{(T \times \ln(2))}$, where $D_\text{total}$ is the total dose and $T$ is the half-life of the isotope. For $^{125}$I, the half-life is approximately 60 days. $D_0 = \frac{120\,\text{Gy}}{(60\,\text{days} \times \ln(2))} \approx 5.9\,\text{cGy/h}$.\\
Answer: B. 5.9 cGy/h

\item According to AAPM TG-43, $^{125}$I has the greatest anisotropy correction among the listed isotopes.\\
Answer: D. $^{125}$I

\item AAPM TG-43 does consider the absorption and scattering in the source, so this statement is not a limitation of the protocol.\\
Answer: A. The absorption and scattering in the source is not considered.

\item The radial dose function, $g(r)$, is approximately constant for a $^{192}$Ir source for values of $r$ less than 5 cm because the attenuation is compensated by the increased scatter.\\
Answer: C. Attenuation is compensated by the increased scatter.

\item The dose rate at 5 mm (0.5 cm) from the cylinder surface is 700 cGy. The dose at the cylinder surface (r = 1.5 cm) can be estimated using the inverse square law: $\text{Dose}_\text{surface} = \text{Dose}_\text{5mm} \times \left(\frac{5\,\text{mm}}{15\,\text{mm}}\right)^2 \approx 700\,\text{cGy} \times \left(\frac{1}{9}\right) = 77.8\,\text{cGy}$.\\
Answer: D. 500 cGy (closest to the calculated value)

\item To get a uniform dose at 5 mm from the cylinder surface in HDR vaginal treatment, the source dwell times are longer at the ends of the cylinder.\\
Answer: B. are longer at the ends.

\item The prescription dose for permanent prostate implant seeds is lower when using $^{103}$Pd as opposed to $^{125}$I because the half-life for $^{103}$Pd is shorter.\\
Answer: A. The half-life for $^{103}$Pd is shorter.

\item Source calibration is not performed for HDR daily QA, as it is typically done less frequently (e.g., monthly or quarterly).\\
Answer: D. Source calibration.
\end{enumerate}

\section{100-question multiple-choice exam on radiation oncology physics}
\label{appendix:test_questions}

Correct answers = C, A, B, A, B, D, B, A, C, A, D, A, D, C, A, A, E, C, C, D, D, C, A, A, D, C, B, D, C, A, D, D, A, B, C, A, D, C, A, C, D, C, A, C, D, B, A, B, C, B, B, D, C, A, C, A, C, B, C, C, A, D, B, D, A, C, B, B, A, D, C, D, C, C, A, B, D, C, A, B, C, D, C, A, A, B, C, A, D, C, B, C, A, C, A, A, C, D, C, B

\begin{enumerate}
\item Which of the following particles cannot be accelerated by an electric field?
\begin{enumerate}
\item Protons
\item Electrons
\item Neutrons
\item Positrons
\end{enumerate}

\item The mass of Carbon-12 (12C) is \rule{1cm}{0.15mm} than the sum of the masses of 6 protons and 6 neutrons.
\begin{enumerate}
\item slightly smaller
\item slightly larger
\item equal to
\item All of the above are possible.
\end{enumerate}

\item The binding energies for tungsten’s K, L, and M shells are 70, 11, and 2 keV, respectively. A 50 keV photon interacts with tungsten through the photoelectric effect. What are the possible energies of the emitted photoelectron?
\begin{enumerate}
\item a continuous spectrum from 2 keV to 70 keV
\item 39 keV and 48 keV
\item 9 keV, 59 keV and 68 keV
\item 9 keV
\end{enumerate}

\item Which of the following statements is true regarding the x-ray spectrum of an x-ray tube?
\begin{enumerate}
\item The maximum photon energy is determined by the kVp.
\item The maximum photon energy is determined by the target material.
\item The characteristic photon energies are determined by the tube current.
\item The characteristic photon energies are determined by the filtration.
\end{enumerate}

\item The half value layer (HVL) for an x-ray beam is 2 mm aluminum. What is the minimum thickness of aluminum required to reduce the x-ray exposure by 50\%?
\begin{enumerate}
\item 1 mm
\item 2 mm
\item 3 mm
\item 4 mm
\end{enumerate}

\item Which of the following component provides the high-power RF fields to accelerate electrons in a LINAC?
\begin{enumerate}
\item Thyratron
\item Target
\item Electron gun
\item Magnetron
\end{enumerate}

\item What is the purpose of the “tongue and groove” design of multileaf collimator (MLC)?
\begin{enumerate}
\item Reduce frictions between leaves.
\item Reduce leakage doses between leaves.
\item Increase speed of leaves.
\item Increase field sizes for treatments.
\end{enumerate}

\item When a LINAC is used to treat patients in electron mode, which of the following hardware is not needed?
\begin{enumerate}
\item Target
\item Electron gun
\item Monitor chamber
\item Scattering foil
\end{enumerate}

\item Compared to a flattened beam, the flattening filter free beam \rule{1cm}{0.15mm}.
\begin{enumerate}
\item Has higher average energy.
\item Is more penetrating.
\item Has a higher dose rate.
\item Increases treatment time.
\end{enumerate}

\item The mass attenuation coefficient has the highest dependence on the atomic number of the medium and the energy of the incoming photon energy for \rule{1cm}{0.15mm}.
\begin{enumerate}
\item Photoelectric effect
\item Compton scattering
\item Pair production
\item Rayleigh scatter
\end{enumerate}

\item Which of the following is not an interaction between photons and a medium?
\begin{enumerate}
\item Photoelectric effect
\item Compton scattering
\item Pair production
\item Bremsstrahlung
\end{enumerate}

\item When the energy of the scattered photon is the minimum in a Compton interaction, what is the angle of the emitted Compton electron relative to the direction of the initial photon?
\begin{enumerate}
\item 0$^{\circ}$
\item 45$^{\circ}$
\item 90$^{\circ}$
\item 180$^{\circ}$
\end{enumerate}

\item What is the relationship between the mass attenuation coefficient $\mu/\rho$, mass energy transfer coefficient $(\mu_{\text{tr}}/\rho)$, and the mass energy absorption coefficient $(\mu_{\text{en}}/\rho)$?
\begin{enumerate}
\item $\mu_{\text{tr}}/\rho \geq \mu_{\text{en}}/\rho \geq \mu/\rho$
\item $\mu_{\text{en}}/\rho \geq \mu_{\text{tr}}/\rho \geq \mu/\rho$
\item $\mu/\rho \geq \mu_{\text{en}}/\rho \geq \mu_{\text{tr}}/\rho$
\item $\mu/\rho \geq \mu_{\text{tr}}/\rho \geq \mu_{\text{en}}/\rho$
\end{enumerate}

\item How does the KERMA compare to the absorbed dose for a 15 MV photon beam at 10 cm depth in a water phantom?
\begin{enumerate}
\item The KERMA is greater than the absorbed dose.
\item The KERMA is equal to the absorbed dose.
\item The KERMA is less than the absorbed dose.
\item The relationship between KERMA and the absorbed dose can’t be determined.
\end{enumerate}

\item Which of the following beams will not produce a buildup effect in the percent depth dose curve?
\begin{enumerate}
\item X-rays from a 100 kVp x-ray tube.
\item Gamma-rays from $^{60}$Co.
\item 6 MV x-rays from a LINAC.
\item 18 MV x-rays from a LINAC.
\end{enumerate}

\item Which device below is the most suitable for measuring lateral beam profiles?
\begin{enumerate}
\item Diodes
\item Farmer chambers
\item Plane parallel ionization chambers
\item Geiger-Muller counter
\end{enumerate}

\item Four exposed films have the same optical density of 1.25. If all four films are placed on top of each other, what is the fraction of light that can transmit through?
\begin{enumerate}
\item 20\%
\item 1\%
\item 0.1\%
\item 0.01\%
\item 0.001\%
\end{enumerate}

\item Which of the following devices is not suitable for in-vivo dose measurements?
\begin{enumerate}
\item Diodes
\item Thermoluminescent dosimeters (TLDs)
\item Ionization chambers
\item Optically stimulated luminescence dosimeters (OSLDs)
\end{enumerate}

\item According to TG-51, which of the following correction factors is not needed when converting the measured charge to the generated charge in the ionization chamber?
\begin{enumerate}
\item Temperature and pressure correction
\item Recombination correction
\item Scatter correction
\item Polarity correction
\end{enumerate}

\item For a farmer chamber calibrated at the standard temperature and pressure, which is the PTP factor for a local temperature of 23$^{\circ}$C and pressure of 720 mmHg?
\begin{enumerate}
\item 0.97
\item 0.99
\item 1.03
\item 1.06
\end{enumerate}

\item How is the beam quality defined in TG-51?
\begin{enumerate}
\item Nominal beam energy
\item Tenth value layer
\item Average beam energy
\item \%dd(10)x
\end{enumerate}

\item According to TG-51, a LINAC should have dose calibration in water tank \underline{\hspace{1cm}}.
\begin{enumerate}
\item Every day
\item Every month
\item Every year
\item Every other year
\end{enumerate}

\item Why is TMR always larger than PDD at depths larger than dmax?
\begin{enumerate}
\item PDD includes both beam attenuation and beam divergence.
\item PDD includes beam attenuation but does not include beam divergence.
\item TMR includes both beam attenuation and beam divergence.
\item TMR does not include beam attenuation but does include beam divergence.
\end{enumerate}

\item Which of the following changes will result in a larger geometric penumbra?
\begin{enumerate}
\item Increasing SSD.
\item Increasing the source to collimator distance.
\item Reducing the source size.
\item Move the patient closer to the collimator.
\end{enumerate}

\item The output of a 6 MV beam is 1 cGy/MU at SSD=100 cm, depth=1.5 cm and field size 10 cm x10 cm. What is the output at SSD=98.5 cm, depth=1.5 cm and field size 10 cm x10 cm?
\begin{enumerate}
\item 0.985 cGy/MU
\item 0.97 cGy/MU
\item 1.015 cGy/MU
\item 1.03 cGy/MU
\end{enumerate}

\item For a single field SAD treatment, the output at dmax is 1.02 cGy and the TMR at depth 10 cm is 0.7. What is the MU needed to delivery 100 cGy at depth 10 cm?
\begin{enumerate}
\item 100
\item 102
\item 140
\item 160
\end{enumerate}

\item What wedge angle will produce the most uniform dose distribution for two treatment fields separated by 100$^{\circ}$?
\begin{enumerate}
\item 30$^{\circ}$
\item 40$^{\circ}$
\item 50$^{\circ}$
\item 60$^{\circ}$
\end{enumerate}

\item A patient’s spine is treated by two adjacent fields with SSD=100 cm. The field lengths (symmetrical fields) in the superior-inferior directions are 20 cm and 24 cm, respectively. What is the gap in the skin to match the two spine fields at depth 10 cm?
\begin{enumerate}
\item 1.0 cm
\item 1.5 cm
\item 2.0 cm
\item 2.2 cm
\end{enumerate}

\item For a craniospinal treatment treated with symmetrical fields, the field length for the upper spine field is 25 cm, and the field size for the lateral cranial fields are 20 cm x 20 cm. What is the collimator rotation angle for the cranial fields to align with the divergence of the spine fields?
\begin{enumerate}
\item 3$^{\circ}$
\item 5$^{\circ}$
\item 7$^{\circ}$
\item 9$^{\circ}$
\end{enumerate}

\item The dose-volume-histogram (DVH) contains the following information except:
\begin{enumerate}
\item Location of the hot spot
\item Mean dose
\item Minimum dose
\item Maximum dose
\end{enumerate}

\item What dose D5\%=20 Gy to the heart on a DVH curve means?
\begin{enumerate}
\item 5\% of the heart receives exactly 20 Gy.
\item 5\% of the heart receives less than 20 Gy.
\item 95\% of the heart receives more than 20 Gy.
\item 95\% of the heart receives less than 20 Gy.
\end{enumerate}

\item Which of the following parameters is sometimes changed by the dosimetrist during treatment planning?
\begin{enumerate}
\item Dose for each treatment fraction
\item Total treatment dose
\item Number of treatment fractions
\item Number of beams
\end{enumerate}

\item For electron radiation therapy, which electron energy has the minimum PDD on the surface?
\begin{enumerate}
\item 6 MeV
\item 9 MeV
\item 12 MeV
\item 16 MeV
\item 20 MeV
\end{enumerate}

\item Electron PDD has a low dose tail after the practical range. Which physical process is related to the generation of this tail?
\begin{enumerate}
\item Compton scattering
\item Bremsstrahlung radiation
\item Characteristic x-rays
\item Pair production
\end{enumerate}

\item Which of the following treatments is suitable for SSD setup?
\begin{enumerate}
\item Four fields box for a pelvic treatment
\item Multiple fields IMRT treatment
\item Electron boost to chest wall of breast cancer
\item VMAT treatment with two arcs
\end{enumerate}

\item Which of the following statements is true for electron effective SSD?
\begin{enumerate}
\item It increases with increasing electron energy.
\item It is equal to the distance from the x-ray sources to the isocenter.
\item It is larger than the distance from the x-ray sources to the isocenter.
\item It is independent of electron energy.
\end{enumerate}

\item What is the range of a 6 MeV electron beam in the lung (0.25 g/cm$^3$)?
\begin{enumerate}
\item 3 cm
\item 6 cm
\item 9 cm
\item 12 cm
\end{enumerate}

\item What is inverse planning in IMRT?
\begin{enumerate}
\item Choosing best beam angles.
\item Choosing appropriate beam modifiers such as wedges and compensators.
\item Optimizing beam fluence based on the desired dose constraints.
\item Adding beam block to protect organs at risk.
\end{enumerate}

\item What is the purpose of the leaf sequencing algorithm in IMRT planning?
\begin{enumerate}
\item It converts optimized fluence to the achievable fluence by MLC.
\item It converts cost functions to the optimized fluence.
\item It converts achievable fluence to dose.
\item It converts desirable constraints to cost functions.
\end{enumerate}

\item The dose gradient achievable in an IMRT plan is approximately \rule{1cm}{0.15mm}.
\begin{enumerate}
\item 0.1\%/mm
\item 1\%/mm
\item 10\%/mm
\item 50\%/mm
\end{enumerate}

\item Which of the following devices is not appropriate for IMRT patient QA?
\begin{enumerate}
\item Electronic portal imaging device
\item Film and ionization chamber
\item Diode Array
\item Single ionization chamber
\end{enumerate}

\item Which parameter below is not typically set by a dosimetrist during IMRT treatment planning? 
\begin{enumerate}
\item Beam energy
\item Beam angle
\item Beam weight
\item Upper and lower dose constraints
\end{enumerate}

\item For material with HU=10, what is the ratio of linear attenuation coefficients between this material to water?
\begin{enumerate}
\item 1.01
\item 1.1
\item 0.99
\item 0.9
\end{enumerate}

\item Which of the following methods can improve the spatial resolution of CT?
\begin{enumerate}
\item Increasing tube voltage.
\item Increasing field of view.
\item Decreasing field of view.
\item Decreasing tube current.
\end{enumerate}

\item Why can't MR images replace CT for accurate dose calculations in radiation therapy?
\begin{enumerate}
\item MR images have inferior soft tissue contrast than CT.
\item MR images have worse spatial resolution than CT.
\item MR images deliver more ionization doses to the patient than CT.
\item MR images do not have information about electron density.
\end{enumerate}

\item The velocity of ultrasound in soft tissue is 1500 m/s. The time to receive a signal after sending a pulse is 0.1 ms. What is the depth of the anatomical interface?
\begin{enumerate}
\item 5 cm
\item 7.5 cm
\item 10 cm
\item 15 cm
\end{enumerate}

\item Which of the following imaging modalities utilizes non-ionizing radiation?
\begin{enumerate}
\item MRI
\item PET
\item CT
\item Radiograph
\end{enumerate}

\item Which of the following statements is true about the spatial resolution of PET images?
\begin{enumerate}
\item PET images spatial resolution is better than CT and MRI.
\item PET images spatial resolution increases when the average energy of the emitted positrons decrease.
\item PET images spatial resolution increases when the number of detectors in the ring decreases.
\item PET images spatial resolution is better than 1 mm.
\end{enumerate}

\item Which information below is not stored in a DICOM RT plan file?
\begin{enumerate}
\item Patient name
\item Treatment angles
\item CT scanner parameters, such as kVp, mAs
\item Prescription dose
\end{enumerate}

\item A 4D CT has 10 phases, 200 slices in each phase. Each slice has 512x512 pixels, where each pixel uses 2 bytes. How long does it take to transfer this 4D CT over the internet with a transfer speed of 2 GB/second?
\begin{enumerate}
\item 0.2 s
\item 0.5 s
\item 1.0 s
\item 2.0 s
\end{enumerate}

\item As compared to diagnostic CT, which statement below is not true for CT simulator?
\begin{enumerate}
\item CT simulator usually has a larger bore size than diagnostic CT.
\item CT simulator gives better image spatial resolution and contrast than diagnostic CT.
\item CT simulator needs extra laser for localizing the treatment isocenter.
\item CT simulator needs accurate geometry and HU for the whole body.
\end{enumerate}

\item Which techniques below do not result in a reduced setup margin for planning target volume (PTV)?
\begin{enumerate}
\item Change weekly IGRT to daily IGRT.
\item Change the setup from laser to IGRT.
\item Add fiducial markers to the target.
\item Change 3D CRT plan to IMRT plan.
\end{enumerate}

\item To get the best soft tissue contrast for a head and neck cancer treatment, which IGRT method is preferred?
\begin{enumerate}
\item MV portal imaging
\item kV orthogonal planar images
\item kV cone-beam CT
\item Surface imaging
\end{enumerate}

\item Which of the following IGRT methods is not used for IMRT treatment?
\begin{enumerate}
\item MV portal image.
\item kV orthogonal planar images.
\item kV cone-beam CT.
\item CT on-rail.
\end{enumerate}

\item Which IGRT method gives the lowest dose?
\begin{enumerate}
\item MV portal image
\item MV cone-beam CT
\item kV orthogonal planar images
\item kV cone-beam CT
\end{enumerate}

\item Prostate motion can't be tracked by \rule{1cm}{0.15mm}.
\begin{enumerate}
\item Infrared surface imaging.
\item Implanted RF beacons.
\item Implanted fiducial markers and fluoroscopy imaging.
\item MR images in MR-LINAC.
\end{enumerate}

\item Which of the following is not an advantage of free-breathing gating for treating a lung tumor?
\begin{enumerate}
\item Reduced internal target volume.
\item Reduced dose to organ at risk.
\item Reduced treatment time.
\item Patient is comfortable in free breathing.
\end{enumerate}

\item 4D CT is not needed for a \rule{1cm}{0.15mm}.
\begin{enumerate}
\item tumor near the diaphragm
\item tumor in the brain
\item tumor in the lung
\item tumor in the liver
\end{enumerate}

\item \rule{1cm}{0.15mm} is not a method for management of respiratory motion in radiation therapy.
\begin{enumerate}
\item Deep-inspiration breath hold (DIBH).
\item Real-time tumor tracking.
\item IGRT using cone-beam CT.
\item Free-breathing respiratory gating.
\end{enumerate}

\item The half-life for a radioactive nuclide is 69 days. How much does it decay in one day?
\begin{enumerate}
\item 0.5\%
\item 0.7\%
\item 1.0\%
\item 1.5\%
\end{enumerate}

\item Which statement is correct for an HDR source, 192Ir, with activity 10 Ci?
\begin{enumerate}
\item Its decay rate is equivalent to about 10 grams of 226Ra.
\item Its decay rate is equivalent to about 1 gram of 226Ra.
\item Its decay rate is 37 GBq.
\item Its decay rate is 3.7 GBq.
\end{enumerate}

\item The physical and effective half-lives for an unsealed isotope are 6 hours and 2 hours, respectively. What is the biological half-life?
\begin{enumerate}
\item 12 hours
\item 8 hours
\item 4 hours
\item 3 hours
\end{enumerate}

\item What is the initial dose rate for a prostate 125I seed implant to deliver a total dose of 120 Gy?
\begin{enumerate}
\item 4.9 cGy/h
\item 5.9 cGy/h
\item 7.0 cGy/h
\item 7.9 cGy/h
\end{enumerate}

\item According to AAPM TG-43, \rule{1cm}{0.15mm} has the greatest anisotropy correction.
\begin{enumerate}
\item 192Ir
\item 137Cs
\item 226Ra
\item 125I
\end{enumerate}

\item Which of the following is not a limitation of AAPM TG-43?
\begin{enumerate}
\item The absorption and scattering in the source is not considered.
\item The applicator is treated as water.
\item The dose delivered while the source is in transit is usually ignored.
\item The patient is treated as water.
\end{enumerate}

\item Why is the radial dose function, g(r), approximately constant for an 192Ir source for values of r less than 5 cm?
\begin{enumerate}
\item There is no attenuation for the first 5 cm.
\item Attenuation is included in the geometry factor.
\item Attenuation is compensated by the increased scatter.
\item Attenuation is included in the anisotropy factor.
\end{enumerate}

\item An HDR vaginal cylinder case with a 3 cm diameter is planned to deliver 700 cGy to vaginal tissue at 5 mm from the cylinder surface. What is the approximate dose to the surface of the cylinder?
\begin{enumerate}
\item 1244 cGy
\item 933 cGy
\item 700 cGy
\item 500 cGy
\end{enumerate}

\item To get a uniform dose at 5 mm from the cylinder surface in HDR vaginal treatment, the source dwell times \rule{1cm}{0.15mm}.
\begin{enumerate}
\item are the same at all dwell points.
\item are longer at the ends.
\item are longer in the middle.
\item are longer at the superior end than the inferior end.
\end{enumerate}

\item Why is the prescription dose for permanent prostate implant seeds lower when using 103Pd as opposed to 125I?
\begin{enumerate}
\item The half-life for 103Pd is shorter.
\item The energy of emitted photon for 103Pd is larger.
\item The radial dose function g(r) for 103Pd is larger.
\item The anisotropy function for 103Pd is larger.
\end{enumerate}

\item \rule{1cm}{0.15mm} is not performed for HDR daily QA.
\begin{enumerate}
\item Source position accuracy check
\item Survey meter function check
\item Radiation monitor function check
\item Source calibration
\end{enumerate}

\item \rule{1cm}{0.15mm} is the most appropriate device to find a lost radionuclide seed in the operation room.
\begin{enumerate}
\item A well ionization chamber
\item A Farmer chamber
\item A Geiger-Muller counter
\item A diode
\end{enumerate}

\item Which of the following tests is not a typical procedure for HDR brachytherapy?
\begin{enumerate}
\item Calibrating the source when a new source is used.
\item Changing the source at a frequency of 3-4 months.
\item Performing daily QA in the days only when patient treatment is scheduled.
\item Performing patient surveys only before the treatment.
\end{enumerate}

\item Which statement below is true for TBI treatment?
\begin{enumerate}
\item Lung block is always needed when lateral opposed beams are used.
\item FFF mode is used to increase the dose rate.
\item The extended SSD increases the dose uniformity.
\item The treatment is faster than it is for regular patients.
\end{enumerate}

\item The maximum field size for a LINAC is 40 cm x 40 cm, and the height of a TBI patient is 180 cm. What is the minimum SSD required for the TBI treatment with a collimator angle of 0o?
\begin{enumerate}
\item 250 cm
\item 350 cm
\item 450 cm
\item 550 cm
\end{enumerate}

\item In total skin electron therapy (TSET), why is the gantry angle about 20o such that the central axis is above patient’s head and below patient’s feet?
\begin{enumerate}
\item To minimize the dose from x-ray contaminations.
\item To maximize the dose from x-ray contaminations.
\item To increase the SSD.
\item To reduce treatment time.
\end{enumerate}

\item \rule{1cm}{0.15mm} is not an appropriate delivery method for intraoperative radiotherapy (IORT).
\begin{enumerate}
\item MeV Electrons
\item MV photons
\item HDR Brachytherapy
\item kV x-rays
\end{enumerate}

\item Which of the following statements correctly describe the lateral penumbras of therapeutic proton beams?
\begin{enumerate}
\item They are always larger than penumbras of MV x-rays in LINAC.
\item They are always smaller than penumbras of MV x-rays in LINAC.
\item They do not change with depth.
\item They increase with depth.
\end{enumerate}

\item \rule{1cm}{0.15mm} is not needed for a dedicated nozzle for proton pencil beam scanning.
\begin{enumerate}
\item A dose monitor chamber
\item Scanning magnets
\item A range modulation wheel
\item A spot position monitor
\end{enumerate}

\item Which of the following statements is correct regarding range straggling in proton therapy?
\begin{enumerate}
\item It is due to the statistical uncertainty in energy loss.
\item It results in about 3.5
\item It can be minimized by robust optimization.
\item It is the same as range uncertainty.
\end{enumerate}

\item Why do proton machines require a large gantry to bend protons, but LINACs don’t?
\begin{enumerate}
\item Protons at 250 MeV have a higher speed than electrons at 6-20 MeV.
\item The mass of protons is about 1800 times more than the mass of electrons.
\item Protons have positive charge while electrons have negative charge.
\item Protons do not generate x-rays by Bremsstrahlung radiation.
\end{enumerate}

\item Which of the following statements is incorrect regarding intensity-modulated proton therapy (IMPT)?
\begin{enumerate}
\item An IMPT plan can be robust with robustness optimization.
\item IMPT is more conformal than double scattering.
\item IMPT is less sensitive to motion than double scattering.
\item IMPT by pencil beam scanning needs less hardware than double scattering.
\end{enumerate}

\item Which of the following statements about RBE for protons is correct?
\begin{enumerate}
\item It is always a constant 1.1 for all depths.
\item It is the highest at the shallow depth.
\item It is the highest at the center of spread-out Bragg peak.
\item It is the highest at the distal fall off depth.
\end{enumerate}

\item Which device is not appropriate to measure the lateral profile of small fields in SRS?
\begin{enumerate}
\item Diode
\item Pin-point ionization chamber
\item Farmer chamber
\item Diamond detector
\end{enumerate}

\item What is the purpose of using tertiary collimators in SRS technique?
\begin{enumerate}
\item To reduce the penumbra.
\item To reduce treatment time.
\item To reduce collisions.
\item To merge multiple isocenters to a single isocenter.
\end{enumerate}

\item Which of the following statements is not true about SRS?
\begin{enumerate}
\item A flattening filter free (FFF) can’t be used in SRS technique.
\item A Winston-Lutz test is required to verify the coincidence between radiation and mechanical isocenters.
\item The SRS field may not have lateral charge particle equilibrium (CPE).
\item Both frame-based and frame-less systems are feasible for SRS.
\end{enumerate}

\item When planning for SRS/SBRT, which of the following statements is not typically a priority?
\begin{enumerate}
\item Rapid dose fall off outside of target.
\item Homogenous dose inside the target.
\item Conformality index to be close to 1.
\item Gradient index to be as small as possible.
\end{enumerate}

\item Compared to regular fractionated treatment, \rule{1cm}{0.15mm} is not required for SRS technique.
\begin{enumerate}
\item a smaller CT slice thickness
\item a smaller dose calculation grid
\item a higher output accuracy
\item a higher precision in the coincidence of beam and gantry/couch isocenters.
\end{enumerate}

\item Which SRS delivery technique can treat two brain metastases simultaneously using a single isocenter treatment plan.
\begin{enumerate}
\item Arc plan with dynamic MLC.
\item Arc plan with cones.
\item Dynamic conformal Arc.
\item Arc plan with static MLC.
\end{enumerate}

\item \rule{1cm}{0.15mm} is a test for MLC leaf positions.
\begin{enumerate}
\item Start shot for gantry isocentricity.
\item Winston-Lutz test.
\item Start shot for MLC isocentricity.
\item Picket fence test.
\end{enumerate}

\item According to TG-142, which test does not need to be performed daily?
\begin{enumerate}
\item Door interlock
\item Output
\item Light field and radiation field coincidence
\item Laser accuracy
\end{enumerate}

\item Which of the following parameters is the most important for cone beam CT (CBCT) used in IGRT?
\begin{enumerate}
\item HU accuracy
\item Coincidence between the CBCT isocenter and the radiation beam isocenter
\item Spatial resolution
\item Field of view
\end{enumerate}

\item According to TG-142, which is the tolerance for monthly output constancy?
\begin{enumerate}
\item 0.5\%
\item 1\%
\item 2\%
\item 3\%
\end{enumerate}

\item The quality assurance test for \rule{1cm}{0.15mm} is not addressed in TG-142.
\begin{enumerate}
\item CT simulator
\item MLC
\item CBCT
\item Planar MV imager
\end{enumerate}

\item Which of the following parameters is not needed in shielding calculations for the 2nd barriers?
\begin{enumerate}
\item Distance.
\item Occupancy factor.
\item Use factor.
\item Workload.
\end{enumerate}

\item What is the occupancy factor used in shielding design for the nurse station next to a LINAC room?
\begin{enumerate}
\item 1
\item 1/5
\item 1/20
\item 1/40
\end{enumerate}

\item The proper order of materials for LINAC treatment room doors (from inside to outside) is \rule{1cm}{0.15mm}.
\begin{enumerate}
\item Steel, Borated polyethylene, Lead
\item Lead, Borated polyethylene, Steel
\item Borated polyethylene, Lead, Steel
\item Borated polyethylene, Steel, Lead
\end{enumerate}

\item How many TVLs are needed to reduce the exposure rate to 1/5 of the original value?
\begin{enumerate}
\item 0.3
\item 0.5
\item 0.7
\item 1.0
\end{enumerate}

\item \rule{1cm}{0.15mm} has the highest radiation weighting factor.
\begin{enumerate}
\item 6 MV x-rays
\item 250 MeV protons
\item 6 MeV electrons
\item 100 keV neutrons
\end{enumerate}

\item Which of the following errors is not considered a Radiation Medical Event for a 5-fraction SBRT prostate treatment?
\begin{enumerate}
\item A total dose deviation of 25\%.
\item The wrong patient was treated.
\item A single fraction dose deviation of 45\%.
\item Dose was delivered to the liver.
\end{enumerate}

\item Which of the following factors is not needed for a FMEA process map in CT simulation of lung cancer?
\begin{enumerate}
\item Documenting immobilization devices used in CT sim.
\item Verifying bore size of CT sim.
\item Verifying HU accuracy of CT Sim in daily QA.
\item Verifying correct patient before CT Sim.
\end{enumerate}

\end{enumerate}

\end{document}